\documentclass[aps,pre,twocolumn]{revtex4} 
\usepackage{graphicx,amsmath}
\usepackage[amssymb]{SIunits}
\usepackage{color}

\begin{document}

\title{Statistics of turbulent fluctuations in counter-rotating Taylor-Couette flows}
\author{Sander G. Huisman}  
\email{s.g.huisman@gmail.com}
\author{Detlef Lohse}
\email{d.lohse@utwente.nl}
\author{Chao Sun}
\email{c.sun@utwente.nl}
\affiliation{Department of Applied Physics and J. M. Burgers Centre for Fluid Dynamics, University of Twente, P.O. Box 217, 7500 AE Enschede, The Netherlands}

\date{\today}

\begin{abstract} 
The statistics of velocity fluctuations of turbulent Taylor-Couette flow are examined. The rotation rate of the inner and outer cylinder are varied while keeping the Taylor number fixed to $1.49 \times 10^{12}$ ($\mathcal{O}(\text{Re})=10^6$). The azimuthal velocity component of the flow is measured using laser Doppler anemometry (LDA). For each experiment $5\times10^6$ datapoints are acquired and carefully analysed. Using extended self-similarity (ESS) \cite{ben93b} the longitudinal structure function exponents are extracted, and are found to weakly depend on the ratio of the rotation rates. For the case where only the inner cylinder rotates the results are in good agreement with results measured by Lewis and Swinney \cite{lew99} using hot-film anemometry. The power spectra shows clear $-5/3$ scaling for the intermediate angular velocity ratios $-\omega_o/\omega_i \in \{0.6, 0.8, 1.0\}$,  roughly $-5/3$ scaling for $-\omega_o/\omega_i \in \{0.2, 0.3, 0.4, 2.0\}$,  and no clear scaling law can be found for $-\omega_0/\omega_i = 0$ (inner cylinder rotation only); the local scaling exponent of the spectra has a strong frequency dependence. We relate these observations to the shape of the probability density function of the azimuthal velocity and the presence of a neutral line.
\end{abstract}

\maketitle

\begin{figure}[ht]
	\begin{center}
		\includegraphics{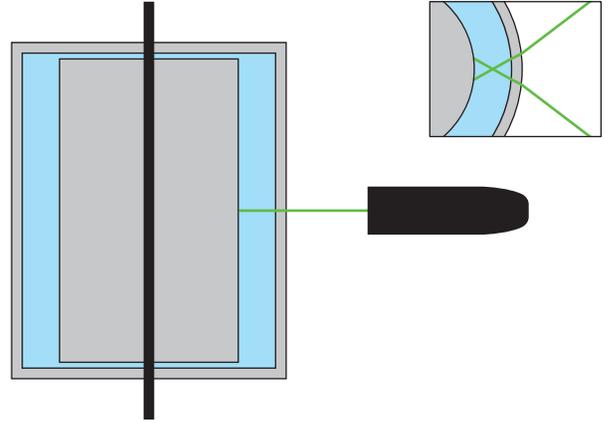}
		\caption{(color online) Sketch of the vertical cross section of the $\text{T}^3\text{C}$ \cite{gil11b}. The beams of the LDA are in the horizontal plane at middle height, $z=L/2$. Top-right inset: Horizontal cross section showing the beams of the LDA. The beams refract twice on the outer cylinder and intersect at the middle of the gap. The angle between the beams are exaggerated to highlight the refraction on the inner and outer surface of the outer cylinder.}
		\label{fig:setup}
	\end{center} 
\end{figure}

Taylor-Couette (TC) flow, among others like Rayleigh-B\'enard convection, and von K\'arm\'an, pipe, channel and plate flow, played a pivatol role in exploring fundamental concepts in fluid mechanics \cite{tay23}. In a TC apparatus, fluid is confined between two independently rotating coaxial cylinders, see fig. \ref{fig:setup}. The TC geometry is best described with cylindrical coordinates: radial distance $\rho$, azimuth $\theta$, and height $z$. The driving of the TC apparatus is given by two Reynolds numbers:
\begin{align*}
 \text{Re}_{i,o}  &= \frac{\omega_{i,o} \rho_{i,o} (\rho_o - \rho_i)}{\nu},
\end{align*}
where $\omega $ is the angular velocity defined as $u_\theta/\rho$, $\rho$ the radius, $\nu$ the kinematic viscosity, and $i$ and $o$ subscripts denote quantities related to the inner and outer cylinder, respectively. Another way of describing the flow is by a Taylor number $\text{Ta} = \frac 14 \sigma (\rho_o - \rho_i)^2 (\rho_o + \rho_i)^2 (\omega_i - \omega_o)^2/\nu^2$, which is the ratio of centrifugal forces to viscous forces, along with a parameter describing the ratio of the driving velocities, for which we have chosen:
\begin{align}
 a &=-\frac{\omega_o}{\omega_i}. \label{eq:a}
\end{align}
$\sigma$ is defined as $((1+\eta)/\sqrt{4\eta})^4$ with the radius ratio $\eta=\rho_i/\rho_o$. By measuring the torque $\mathcal{T}$ \cite{lew99,dub05,ji06,gil11,	pao11,hui12,merbold2013,brauckmann2013}, required to maintain constant angular velocity of both cylinders, we can find the power input ($P$) of our system using $P =  \mathcal{T} \left|\omega_i - \omega_o \right|$. Note that we can measure the torque on either cylinder as it has the same magnitude on the inner and the outer cylinder\cite{egl2007}. As all the energy that enters the system globally will be dissipated by viscous dissipation, the torque can be related to the average energy dissipate rate:
\begin{align}
 \epsilon &= \frac{\text{power input}}{\text{mass}} = \frac{\mathcal{T} \left|\omega_i - \omega_o \right|}{\rho_\text{fluid} \pi (\rho_o^2 - \rho_i^2) L}, \label{eq:dissip}
\end{align}
where $\rho_\text{fluid}$ is the density of the working fluid, and $L$ the length of the cylinders. Using the energy dissipation rate and the viscosity we can now find the average Kolmogorov length scale \cite{kol41,kol41b} in our flow: $\eta_K=\left( \nu^3 / \epsilon  \right)^{1/4}$. 

Using hot-film anemometry, Lewis and Swinney \cite{lew99,she01} measured the statistics of velocity fluctuations for the case of inner cylinder rotation for $\text{Re}$ up to $5.4 \times 10^5$. They found that the energy spectrum does not show power law scaling, and that the structure function exponents---calculated using extended self-similarity \cite{ben93b}---are close to those found in other flows \cite{lew99}. In this paper we quantify the turbulent statistics of Taylor-Couette flow with various rotation ratios (eq. \ref{eq:a}) for fixed Taylor number of $\text{Ta}=1.49\times 10^{12}$ ($\mathcal{O}(\text{Re})=10^6$). We used the Twente Turbulent Taylor-Couette facility ($\text{T}^3\text{C}$) \cite{gil11b}, which was filled with water and actively cooled to keep the temperature constant. The $\text{T}^3\text{C}$ has an inner cylinder with an outer radius of $\rho_i = \unit{200}{\milli \meter}$, a transparent outer cylinder with inner radius $\rho_o = \unit{279}{\milli \meter}$, giving a radius ratio of $\eta = 0.716$. The cylinders have a height of $L = \unit{927}{\milli \meter}$, resulting in an aspect ratio of $\Gamma = L/(\rho_o - \rho_i) = 11.7$. We measured the azimuthal velocity using laser Doppler anemometry (LDA). The advantage of this technique is that it allows for a non-intrusive measurement of a velocity component. For the case of counter-rotation the mean flow direction is not always in a single direction, using \textit{e.g.}~a hot film probe or a pitot tube to measure the local velocity would result in measuring the \textit{speed} in the wake of the probe. The laser beams go through the outer cylinder and are focused in the middle of the gap ($2\rho=\rho_i+\rho_o$) at mid height ($z=L/2$), and lie in the $\theta$--$r$ plane, see also figure \ref{fig:setup}. The water is seeded with tracer particles \cite{dantec} with a mean radius of \unit{2.5}{\micro \meter} and density of \unit{1.03}{\gram \per \centi \meter^3}. This radius is roughly 6 times smaller than our Kolmogorov length scale. We equate the drag force $F_\text{drag}=6\pi \mu r_\text{seed} \Delta u_\theta$ and the centrifugal force $F_\text{cent}=(\rho_\text{seed}-\rho_\text{fluid}) (4\pi r^3/3) u_\theta^2 / \rho$ of the seeding particle, and compute $\Delta u_\theta=|u_{\theta,\text{seed}}-u_{\theta,\text{fluid}}|$ to make sure the particles faithfully follow the flow. For our measurements we find that $\Delta u_\theta \approx \unit{40}{\micro \meter \per \second}$ which is much smaller than the driving velocities $\mathcal{O}(\omega_i \rho_i-\omega_o \rho_o) = \unit{10}{\meter \per \second}$, so we are sure that the particles follow the flow. Due to the curvature of the outer cylinder we have to correct the measured velocity by multiplying it with a constant factor. We find this constant numerically by ray-tracing \cite{hui12lda} the LDA beams in our optical geometry.

\begin{table}[ht]
  \centering
 		\begin{tabular}{ccccccccc}
$a$ & $\frac{\omega _i}{2\pi}$ & $\frac{\omega _o}{2\pi}$ & $\text{Ta}$ & $\text{Re}$ & $\text{Re}_\lambda$ & $ \left \langle u_{\theta} \right \rangle_t$  & $\sigma$ & $\frac{\mu_4}{\sigma^4}$ \\ [1mm]
& \unit{}{1/s} & \unit{}{1/s} & $10^{12}$ & $10^6$ & & \unit{}{m/s} & \unit{}{m/s} & \\ [1mm]
\hline 
 0.0 & 9.99 & 0.00 & 1.49 & 1.38 & 106 & 5.08 & 0.31 & 3.37 \\
 0.2 & 8.32 & -1.66 & 1.49 & 1.32 & 144 & 2.42 & 0.38 & 3.00 \\
 0.3 & 7.67 & -2.30 & 1.49 & 1.29 & 97 & 1.62 & 0.32 & 3.96 \\
 0.4 & 7.13 & -2.85 & 1.49 & 1.27 & 120 & 1.48 & 0.35 & 2.77 \\
 0.6 & 6.24 & -3.74 & 1.49 & 1.24 & 240 & 0.36 & 0.47 & 3.02 \\
 0.8 & 5.54 & -4.44 & 1.49 & 1.21 & 278 & -0.43 & 0.49 & 2.89 \\
 1.0 & 4.99 & -5.00 & 1.49 & 1.19 & 235 & -0.83 & 0.43 & 3.27 \\
 2.0 & 3.32 & -6.66 & 1.49 & 1.12 & 173 & -3.10 & 0.32 & 3.23 \\
		\end{tabular}
  \caption{Experimental parameters for the various rotation ratios. For each experiment the Taylor number is fixed and the number of datapoints is also fixed at $5\cdot 10^6$. $\omega_{i}$ and $\omega_o$ are measured using high precision magnetic encoders. All the measurements are done at mid height and in the middle of the gap. $\langle . \rangle_t$ denotes averaging over time. The Taylor-Reynolds number is found by combining local velocity and global torque measurements: $\text{Re}_\lambda=\sqrt{\frac{15\sigma^4}{\epsilon \nu}}$, where $\epsilon$ comes from the global torque, see eq. (\ref{eq:dissip}). The standard deviation of $u_\theta$ is given by $\sigma$ and the kurtosis by $\frac{\mu_4}{\sigma^4}$.}
  \label{table:experiments}
\end{table}

In our experiments we fixed $\Delta \omega \equiv \omega_i -\omega_o$, see table \ref{table:experiments}, the consequence of this is that our Taylor number is fixed, while our Reynolds number varies slightly. For each experiments we acquire 5 million data points. Because the arrival time of LDA measurements are of stochastic nature the time-series are then linearly interpolated using twice the average acquisition frequency, such as to create a time series with equal temporal spacing. We chose double the frequency to also capture fast fluctuations which are otherwise lost. It can happen that for a relatively long time there is no measurement, while for other moments a burst of measurements are taken. We take care of `disabling' parts of the interpolated time-series for which the temporal gap in the original data is too big, and that interpolation is not justified. These disabled data-points are not used in any of the calculations, except for the spectra.

\begin{figure}[ht]
	\begin{center}
		\includegraphics{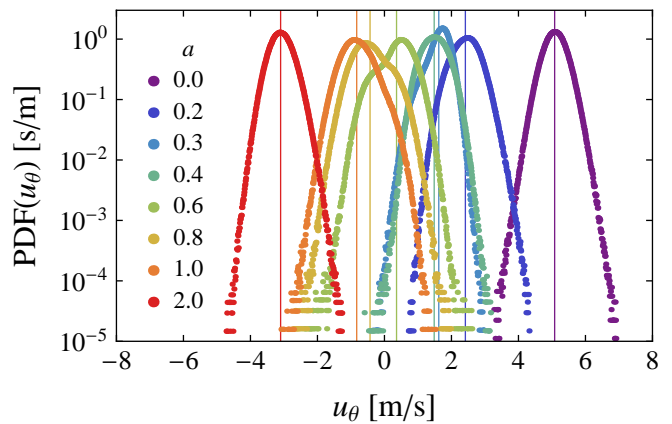}
		\caption{(color online) Probability density functions of the azimuthal velocity ($\text{PDF}(u_\theta)$) measured at mid-height for varying $a=-\omega_o/\omega_i$. The mean of the velocities are indicated by grid lines with their respective color.}
		\label{fig:logpdf}
	\end{center} 
\end{figure}

We will first look at the probability density function (PDF) of the velocimetry data, see fig. \ref{fig:logpdf}. For the cases $a=0.6$, $a=0.8$, and $a=1$, we see that the PDF seems to be comprised of two parts, as it has two bumps; one bump for $u_\theta>0$ and one for $u_\theta<0$. It seems that the presence of a neutral line ($u_\theta=0$) alters the flow dynamics, as was also found in ref. \cite{gils2012jfm}. The outer region is stabilized by the outer cylinder, while the inner region is destabilized by the inner cylinder. Somewhere in between there must therefore be an interface were $u_\theta=0$. For $a \in \{0.6, 0.8, 1.0\}$ our measurement position is on the border of this interface \cite{gils2012jfm}. Also indicated in fig. \ref{fig:logpdf} and in table \ref{table:experiments} is the mean velocity. As expected, the mean velocity decreases monotonically with increasing $a$, while the standard deviation is quite similar throughout, but slightly higher for the cases where the PDF is comprised of two contributions \cite{gils2012jfm}.

\begin{figure}[ht]
	\begin{center}
		\includegraphics{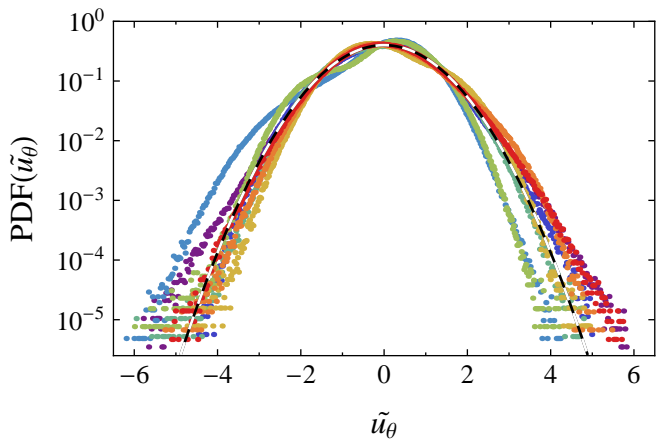}
		\caption{(color online) Probability density function of the standardized azimuthal velocity ($\text{PDF}(\tilde{u_\theta})$) for various $a$. Colors are the same as in fig. \ref{fig:logpdf}. The dashed line is a Gaussian with zero mean and unit variance.}
		\label{fig:standardizedlogpdf}
	\end{center} 
\end{figure}

In order to compare the shape of the PDFs we standardize the velocities (transforming the data such that it has zero mean and unit variance):
\begin{align}
     \mu_n &= \big \langle \big( u_\theta - \left \langle u_\theta \right \rangle_t \big)^n  \big \rangle_t \nonumber \\
     \sigma &= \sqrt{\mu_2} \label{eq:sigma} \\
	 \tilde{u_\theta} &= \left \langle u_\theta -  \left \langle u_\theta \right \rangle_t \right \rangle_t / \sigma \nonumber
\end{align}
where $\langle . \rangle_t$ denotes averaging over time. In fig. \ref{fig:standardizedlogpdf} we plot the PDFs of the standardized velocities. We now see that the tails of most of the distributions behave much like a Gaussian. This is also reflected in the value for the kurtosis ($\mu_4/\sigma^4$), see table \ref{table:experiments}. The values that we find are close to 3 (except for the case $a=0.3$) and the distributions are only slightly leptokurtic or platykurtic.

Table \ref{table:experiments} also includes the Taylor-Reynolds number $\text{Re}_\lambda$. We find that the Taylor-Reynolds number is not necessarily the largest for the case when the torque is the highest ($a\approx 0.33$), as a higher torque means a higher $\sigma$ (eq. \ref{eq:sigma}), but also a higher $\epsilon$ (eq. \ref{eq:dissip}). Furthermore, we note that $\sigma$ depends on the radial and axial position and that therefore the Taylor-Reynolds number is a function of position.

\begin{figure}[ht]
	\begin{center}
		\includegraphics{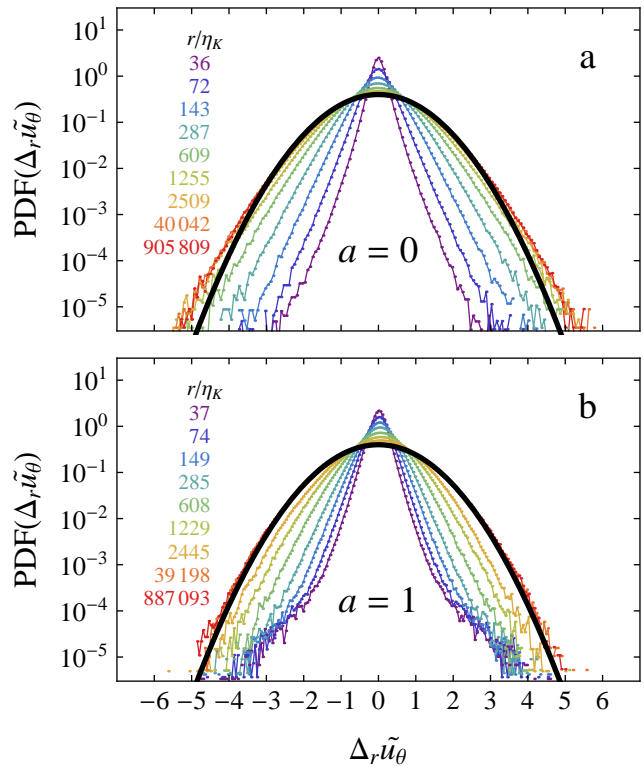}
		\caption{(color online) (a) Probability density functions of velocity increments ($\text{PDF}(\Delta_r u_\theta)$) for varying $r$ for the case of $a=0$. Values of $r$ are shown as multiples of the Kolmogorov length scale, and are colored accordingly. (b) Same as in fig. a but for $a=1$. (a--b) The black line is a Gaussian with zero mean and unit variance.}
		\label{fig:incremental}
	\end{center} 
\end{figure}

Though the PDF of the velocity is of importance in describing a turbulent flow, it obviously does not describe the dynamics of the flow. We therefore look at velocity increments $\Delta_r u_\theta$:
\begin{align}
  \Delta_r u_\theta &\equiv u_\theta(x+r) - u_\theta(x). \label{eq:velocityincrements}
\end{align}
Here $r$ is the distance between the two measurement positions $x$ and $x+r$. As we only probe the velocity at one position, we have to invoke Taylor's frozen flow hypothesis \cite{fri95} to obtain the velocity increments: $u_\theta(x+r,t) = u_\theta(x,t-r/U)$ where $U$ is a typical velocity scale. Here we chose the rms velocity of $u_\theta$, as it best describes the displacement of a fluid parcel for the cases where the velocity is in both directions ($a=0.6$, $a=0.8$, and $a=1.0$). For the cases where the velocity is mainly in one direction the rms velocity is very close to the absolute of the mean velocity. We plot the PDF of $\Delta_r u_\theta$ [eq. (\ref{eq:velocityincrements})] for several different $r$ for the cases of $a=0$ and $a=1$, see figures \ref{fig:incremental}.

\begin{figure*}[ht!]
	\centering
	\includegraphics{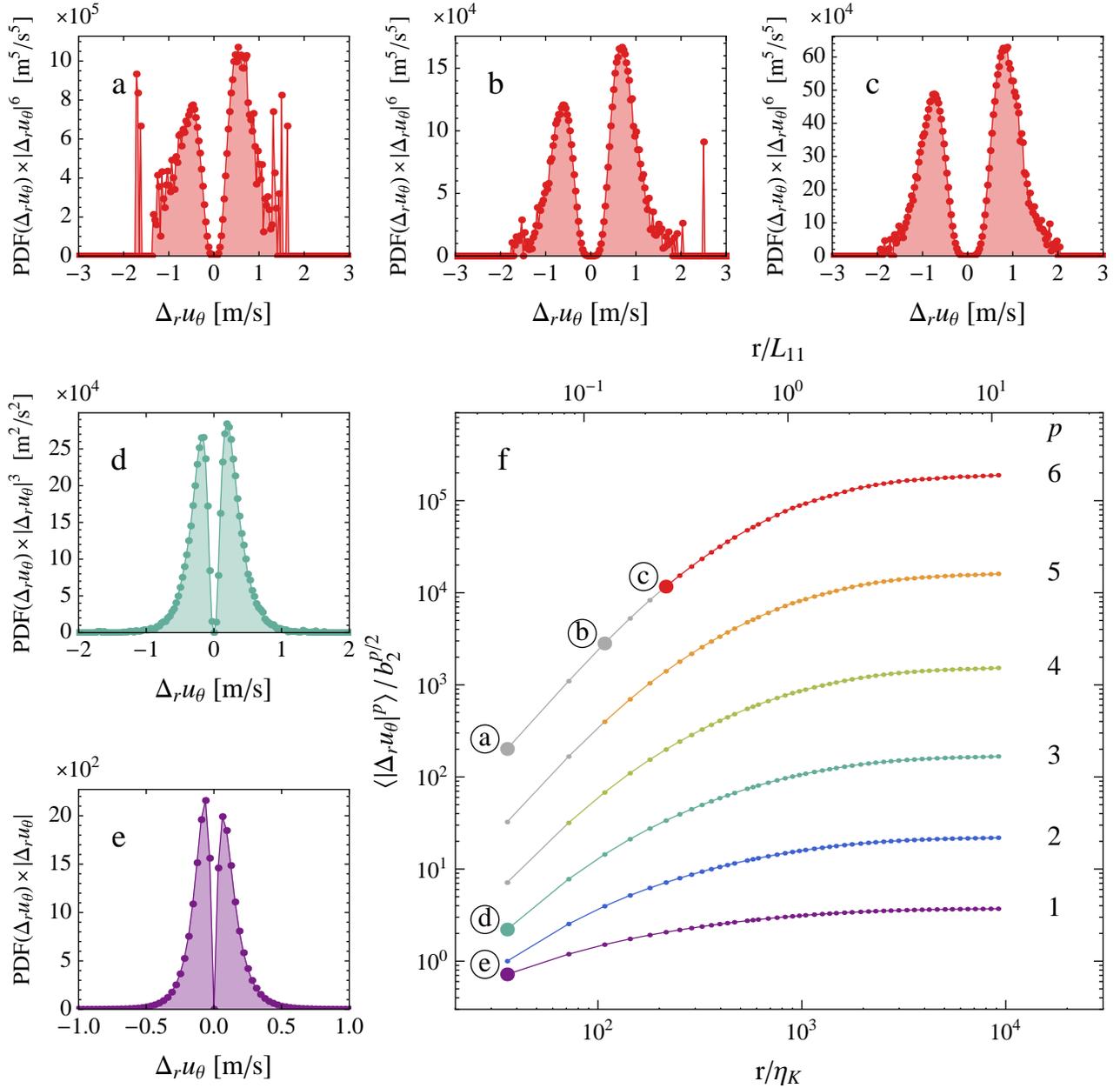}
	\caption{(color online) (f) Longitudinal structure functions ($D^*_p(r)$) for varying powers $p$ (indicated on the right) for the case of $a=0$. Structure functions are scaled using a constant $b_2=D^*_2(r_{\text{min}})$. The separation distance $r$ is normalized using the Kolmogorov length scale $\eta_K$ and the integral length scale $L_{11}$. For various $p$ and $r$ (thick points labeled \textcircled{a}--\textcircled{e}) the integrand of equation (\ref{eq:integratestrucfunc}) is plotted. Data in gray color is non fully converged, and is omitted in the ESS calculation. (a--e) Integrand of eq. (\ref{eq:integratestrucfunc}) for various positions in fig. f. The area under the graph are the values of $D_p^*(r)$. (a,b) Non-converged structure function integrands. (c--e) Converged structure function integrands.}
	\label{fig:multistructfunc}
\end{figure*}

For both $a$ we can clearly see that for small $r/\eta_K$ the distributions are very leptokurtic (the kurtosis $\mu_4/\sigma^4=3.3\times10^3$ for $r/\eta_K=36$, $a=0$, and $\mu_4/\sigma^4=2.4\times10^3$ for $r/\eta_K=37$, $a=1$). Here $\eta_K$ is calculated based on globally measured torque values at the corresponding $a$. Surprisingly, at $a=0$ the PDF does not recover to a Gaussian (Fig. \ref{fig:incremental}a) for very large $r$, whereas it does become normally distributed for other $a$, see \textit{e.g.} fig. \ref{fig:incremental}b for $a=1$. We speculate that for large $r$ and due to the periodic nature of our geometry and the coherent structures in our flow (Taylor vortices), that the flow can stay correlated for an unusually long time for certain $a$ and for certain positions in the flow. We will systematically characterize the $r$ dependence of different moments using longitudinal structure functions: $D_p (r) = \left \langle \left(   \Delta_r u_\theta \right)^p \right \rangle$. For odd moments $p$, $D_p(r)$ is converging very slow, we therefore use the absolute \cite{reeha,reehb} value of the velocity increments:
\begin{align}
	 D_p^* &= \left \langle \left| \Delta_r u_\theta \right|^p \right \rangle \label{eq:strucfunc}
\end{align}
for all $p$. Note that there is no theoretical justification that $D_p^*(r)$ and $D_p(r)$ scale the same way, though the scaling of $D_p^*$  (for odd $p$)  has been found to be similar to that of $D_p$ but not essentially the same, see \textit{e.g.} \cite{Stolovitzky1993,Herweijer1995}. Figure \ref{fig:multistructfunc}f shows the structure functions for $p \in [1,6]$ for the case of $a=0$. We carefully examine the convergence when computing the structure functions. While $D_p^*$ can be calculated using eq. (\ref{eq:strucfunc}), we can alternatively express it as an integral:
\begin{align}
 D_p^* &= \int \limits_{-\infty}^\infty \text{PDF}(\Delta_r u_\theta) |\Delta_r u_\theta|^p \mathrm{d}(\Delta_r u_\theta). \label{eq:integratestrucfunc}
\end{align}

\begin{figure*}[ht!]
	\centering
	\includegraphics{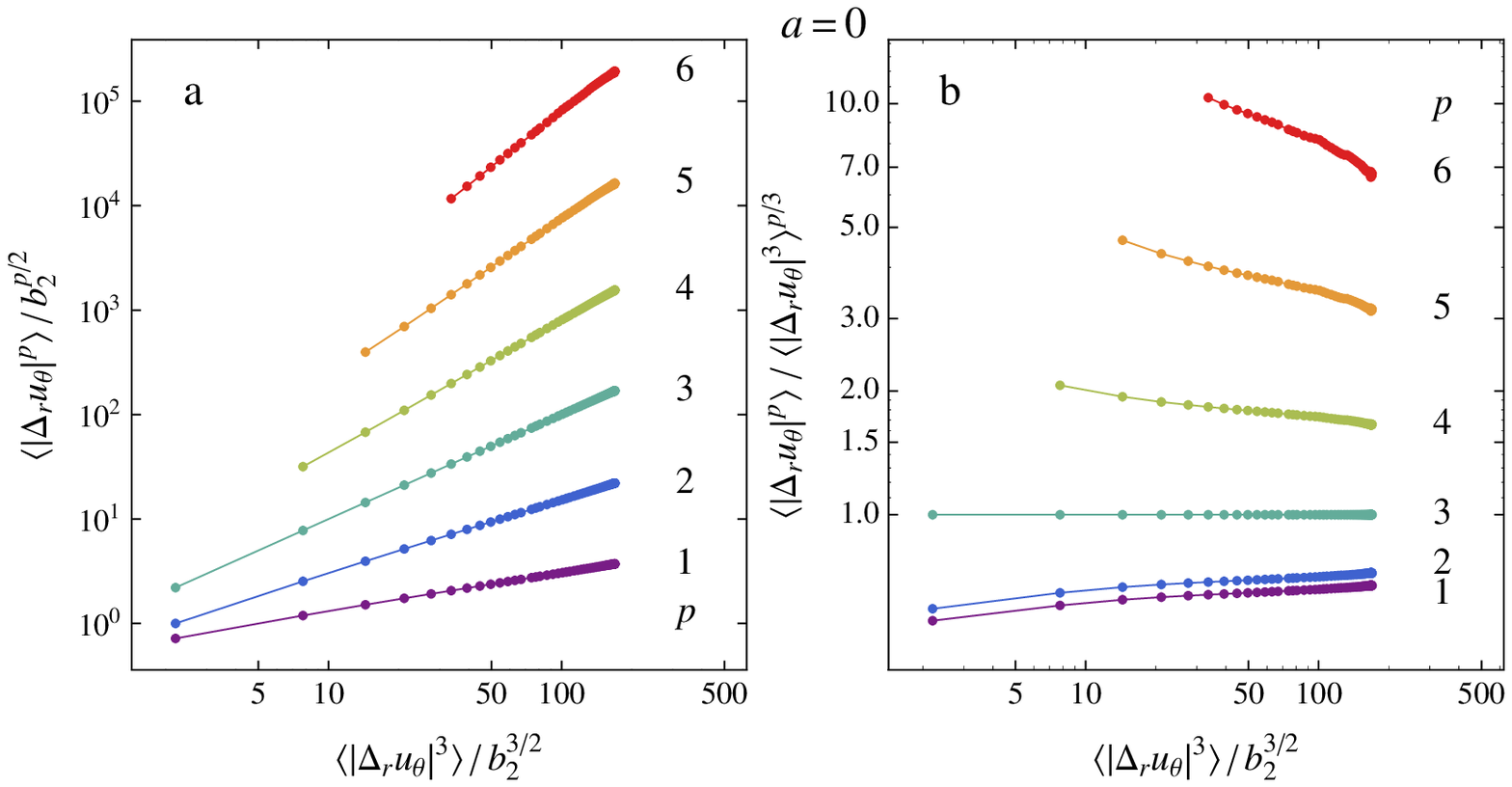}
	\caption{(color online) (a) Extended self similarity for the case of $a=0$. Powers $p$ are indicated on the right. Power law scaling can be seen for each power $p$. (b) Extended self similarity compensated by Kolmogorov scaling $p/3$. The quality of the scaling can now be seen better. Gray data in figure \ref{fig:multistructfunc}f are omitted in this analysis as they are not fully statistically converged.}
	\label{fig:ess}
\end{figure*}

\begin{figure*}[ht!]
	\centering
	\includegraphics{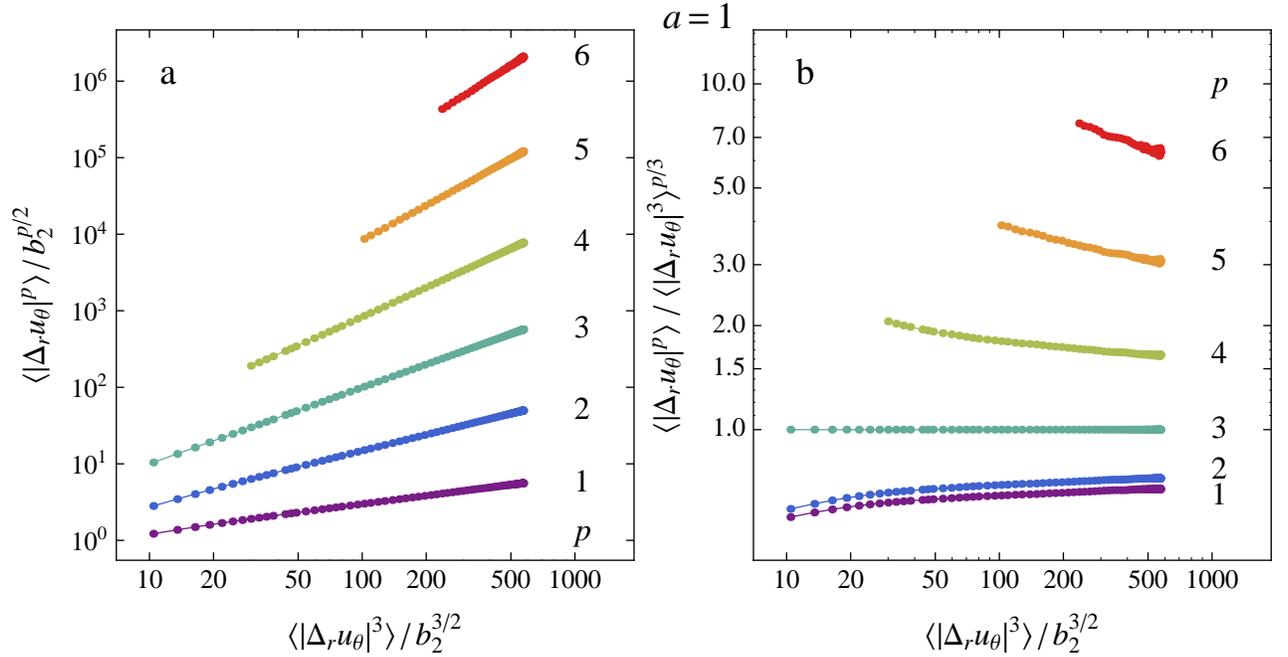}
	\caption{(color online) (a) Extended self similarity for the case of $a=1$. Powers $p$ are indicated on the right. Power law scaling can be seen for each power $p$. (b) Extended self similarity compensated by Kolmogorov scaling $p/3$. The quality of the scaling can now be seen better. Similar to the analysis of figure \ref{fig:multistructfunc} ($a=0$), non-converged data is omitted from the structure functions.}
	\label{fig:ess2}
\end{figure*}

We now take a careful look at the integrand of eq. (\ref{eq:integratestrucfunc}) and plot this integrand for points \textcircled{a}--\textcircled{e} in figure \ref{fig:multistructfunc}f, corresponding to figures \ref{fig:multistructfunc}a--\ref{fig:multistructfunc}e, respectively. For increasing $p$ the structure functions measure the influence of increasingly rare events. We see in figs. \ref{fig:multistructfunc}a and \ref{fig:multistructfunc}b that the tails of the integrands at point \textcircled{a} and \textcircled{b} are not fully converged, and we therefore have to omit these points from the structure functions. We do this analysis for all $p$ and for all $r$ for each case of $a$ and omit data from the analysis; this omitted data is plotted gray in fig. \ref{fig:multistructfunc}f. For high $p$ and low $r$ we find that we do not have sufficient statistics to capture all the rare events necessary to faithfully calculate $D_p^*$.

For fully developed turbulent flows the structure functions are suggested to scale as power laws in the inertial subrange \cite{fri95}:
\begin{align*}
 D_p(r) &\propto r^{\zeta_p}
\end{align*}
Kolmogorov predicted \cite{kol41,kol41b,fri95} these exponents to scale as $\zeta_p = p/3$; any deviation from this model is the result from the intermittency of the flow. For our case we will denote the scaling exponents as $\zeta_p^*$, as we are using $D_p^*(r)$ in the analysis rather than $D_p(r)$. We expect an inertial subrange for roughly $10\eta <r< L_{11}$---where $L_{11}$ is the integral length scale---because these length scales are separated by roughly 2 decades for our flow. But, as seen in figure \ref{fig:multistructfunc}f, we don't see an inertial subrange where the structure functions show power-law scaling, a finding also observed in ref. \cite{lew99}. We therefore are unable to extract structure function exponents directly from our structure functions; this results holds for all our $a$. As suggested by Benzi \textit{et al.} \cite{ben93b}, we can extend the range over which scaling holds by employing extended self-similarity (ESS). The $p$th order structure functions is plotted as a function of the 3rd order structure function; the scaling exponent is now given by $\zeta_p^*/\zeta_3^*$. From the Navier--Stokes equations one can derive the Kolmogorov--Howard--von K\'arm\'an (KHvK) relation, from which we can determine that for $r \gg \nu^{3/4} \epsilon^{-1/4}$, $D_3 \propto -\frac 45 \epsilon r$, or $\zeta_3=1$. In this work we will assume that $\zeta_3 = \zeta_3^* =1$, and we would like to note that the KHvK relation is derived under the assumption of isotropic homogenous turbulence, which is questionable in our flow arrangement. Nevertheless, we employ ESS analysis to our longitudinal structure functions, see figure \ref{fig:ess}a ($a=0$) and figure \ref{fig:ess2}a ($a=1$).

\begin{figure}[ht!]
	\begin{center}
		\includegraphics{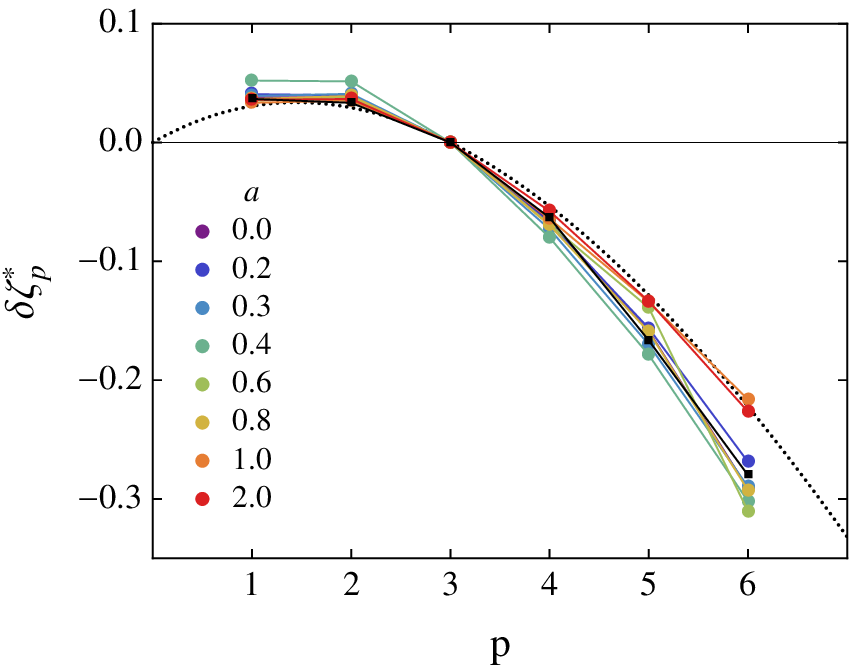}
		\caption{(color online) Longitudinal structure function exponents deviations: $\delta\zeta_p^* = \zeta_p^* -p/3$. Black dotted line shows the She-L\'ev\^eque model ($\zeta_p = p/9 + 2 + 2 (2/3)^{p/3}$). Black squares are the data from Lewis and Swinney \cite{lew99} for $a=0$.}
		\label{fig:compexponentplot}
	\end{center} 
\end{figure}

We now clearly see that in the ESS representation the scaling is much better, and that we are able to extract structure function exponents from our data. In addition, we also compensate our data with Kolmogorov's prediction ($p/3$), see figs. \ref{fig:ess}b and \ref{fig:ess2}b. Note that in these compensated plots \cite{reeha,reehb} (plotted in log-log) any deviation from a perfect power-scaling is amplified and clearly visible. Furthermore, we expect a straight line in the case of perfect scaling. We perform power-law fits to our ESS data for all $a$, and extract $\zeta_p^*$, see table \ref{table:exponents} and figure \ref{fig:compexponentplot}.

\begin{table}[ht]
  \centering
 		\begin{tabular}{p{0.5cm} ccccccccc}
  $a$ & 0 & $0_{\text{LS}}$ & 0.2 & 0.3 & 0.4 & 0.6 & 0.8 & 1 & 2 \\
		 \hline
		 $p$ \\
		 1 & 0.37 & 0.37 & 0.37 & 0.37 & 0.39 & 0.37 & 0.37 & 0.37 & 0.37 \\
		 2 & 0.70 & 0.70 & 0.71 & 0.71 & 0.72 & 0.70 & 0.71 & 0.70 & 0.70 \\
		 $3^*$ & 1 & 1 & 1 & 1 & 1 & 1 & 1 & 1 & 1 \\
		 4 & 1.27 & 1.27 & 1.27 & 1.26 & 1.25 & 1.26 & 1.26 & 1.27 & 1.28 \\
		 5 & 1.51 & 1.50 & 1.51 & 1.50 & 1.49 & 1.53 & 1.51 & 1.53 & 1.53 \\
		 6 & 1.71 & 1.72 & 1.73 & 1.71 & 1.70 & 1.69 & 1.71 & 1.78 & 1.77 \\
		\end{tabular}
  \caption{Structure function exponents $\zeta_p^*$, for different a. $0_{\text{LS}}$ is the data from Lewis \& Swinney \cite{lew99}, for which $a=0$ and $\text{Re}=5.4\times10^5$. Because of the usage of ESS by definition $\zeta_3^* \equiv 1$.}
  \label{table:exponents}
\end{table}

\begin{figure*}[ht!]
	\centering
	\includegraphics{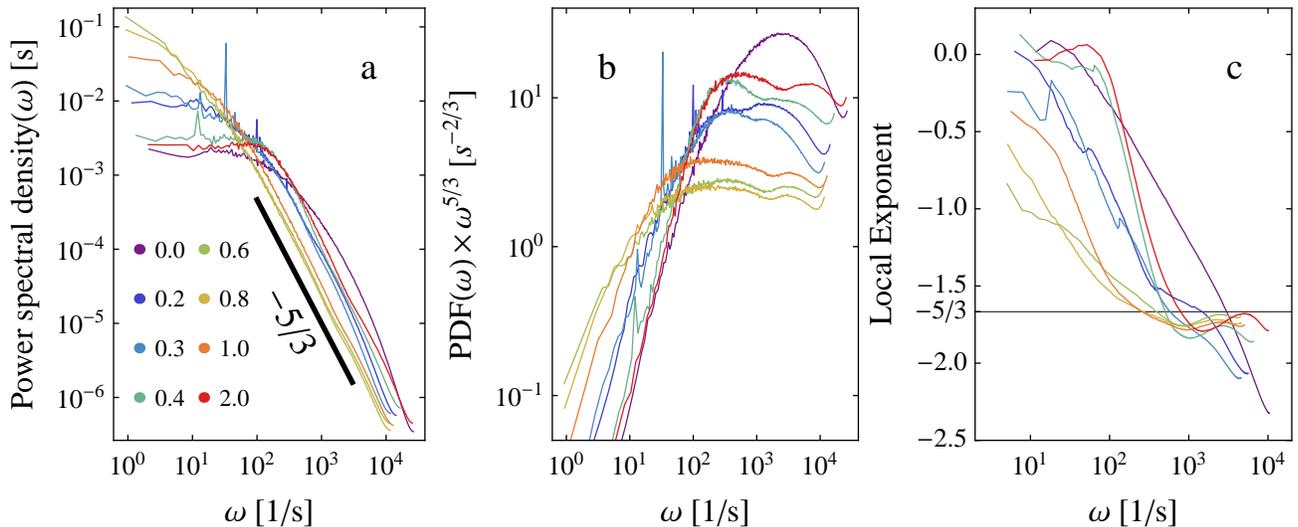}
	\caption{(color online)  (a) Spectral density of the azimuthal velocity for various $a$, $-5/3$ scaling is indicated with a black line. (b) Spectral density compensated by $\omega^{5/3}$. (c) Local scaling exponent over an interval of 1 decade as a function of frequency $\omega$. Kolmogorov's inertial subrange scaling ($-5/3$) is included as a horizontal line. The legend displayed in figure a corresponds also to the colors used in figure b and c.}
	\label{fig:spectra}
\end{figure*}

Lewis and Swinney \cite{lew99} have performed similar analysis to Taylor-Couette flow at pure inner cylinder rotation, and we have included their structure function exponents (obtained using ESS, $\text{Re}=5.4\times10^5$) also in the aforementioned table. We see that for $1 \leq p \leq 5$ the structure function exponents are similar, independent of the Reynolds number, and almost independent of the amount of counter rotation applied to the system. It is only for $p=6$ that the differences between the exponents become noticable; $1.69$ for $a=0.6$ and $1.78$ for $a=0$. This difference might be caused by not including for high $p$ and low $r$-data in our fits due to lack of statistics. We notice that for increased counter-rotation (\textit{i.e.}, increased $a$) we do not see a systematic trend for the structure function exponents. From global torque measurements \cite{gil11,pao11,gils2012jfm,ostilla2013} we know that there is a maximum in the torque needed to sustain a constant angular velocity for both cylinders. This peak in the torque has been found around $a\approx0.33$. This peak in the torque has as a consequence that the Kolmogorov lengthscale is smallest for the rotation ratio $a=0.3$ ($\eta_K = \unit{15.7}{\micro \meter}$) and largest for $a=2.0$ ($\eta_K = \unit{21.0}{\micro \meter}$)---assuming homogenous turbulence for both cases. We, however, don't see any similar trend in the structure function exponents. It seems that all the exponents are nearly \textit{universal}; independent of the Reynolds number (Lewis and Swinney \cite{lew99} also performed their experiments for $\text{Re}=6.9\times10^4$ and found similar structure function exponents as for our $\text{Re}=1.38\times10^6$) and independent on the amount of counter-rotation applied to the system. Our structure function are tabulated in table \ref{table:exponents} and plotted in figure \ref{fig:compexponentplot}, for comparison we have included the data from Lewis and Swinney \cite{lew99} and the She-L\'ev\^eque model with its standard parameters \cite{she1994}. We find that for nearly all the exponents, and for nearly all the $a$ cases that Taylor-Couette flows seems more intermittent than the She-L\'ev\^eque model predicts, as already reported by ref. \cite{lew99}.

In fig. \ref{fig:spectra}a we plot the spectra for all cases. We see that most of the energy is kept in the low frequencies \cite{caballero2013}, even lower than the driving frequency: $\omega_i-\omega_o=\unit{2\pi 10}{\second^{-1}}$ for the cases $a\in\{0.6,0.8,1.0\}$. Furthermore, we see for $a=0$ that we do not have any power-law scaling behavior, as already been found in TC experiments of ref. \cite{lew99}. However for $a\in\{0.6,0.8,1.0\}$---the same $a$s for which the PDF of the azimuthal velocity showed that it was made up of two distributions---we see a power law scaling with exponent $-5/3$. To reveal the quality of the scaling we compensate the data by $\omega^{5/3}$, see figure \ref{fig:spectra}b. We indeed see in figure \ref{fig:spectra}b that the power spectra for $a\in\{0.6,0.8,1.0\}$ levels over roughly two decades. Furthermore, for the case $a=0$ we see that there is no scaling what so ever. In addition, we also plot the local slope of the spectrum. Here we also see that the power-law scaling for $a\in\{0.6,0.8,1.0\}$ is around $-5/3$ around $\omega=\unit{10^3}{\second^{-1}}$, and that for $a=0$ the exponent is constantly changing. The curving-up of our spectra at the high frequency end is due to our limited measuring frequency. We therefore do not recover the steep slopes ($\gg2$) found by Lewis \& Swinney \cite{lew99}. However, they also found that for $a=0$ the local slope is never constant, and is a monotonic function of $\omega$. Of course, we relied on Taylor's hypothesis, which certainly has to be experimentally justified in the future using a field measurement technique, \text{e.g.} particle imaging velocimetry. In addition we assumed concepts of homogeneous isotropic turbulence to obtain $\epsilon$ and therefore $\eta$. Future work will be necessary to study the anisotropic properties of the flow \cite{SO3,procaccia05}.

To summarize, we have measured the local azimuthal velocity in a turbulent Taylor-Couette flow with various amount of counter rotation using laser Doppler anemometry. We found that the structure functions do not show an inertial subrange for all $p$ and for all cases $a$. Using extended self-similarity analysis \cite{ben93b} we extracted the structure function exponents, which are in good agreement with earlier results by Lewis and Swinney \cite{lew99}. We find that for $a=0$ the structure function exponents are nearly independent of the Reynolds number: previous results \cite{lew99} are for $\text{Re}=6.9\times10^4$ and $\text{Re}=5.4\times10^5$, while our current results are for $\text{Re}=1.38\times10^6$. Any discrepancy between these exponent could easily be caused by different fitting intervals and are certainly within experimental error. Furthermore, we find that adding rotation of the outer cylinder of the system to create counter-rotation does not strongly influence the structure function exponents, but does strongly change the scaling of the spectra. While for $a\in\{0.6,0.8,1.0\}$ we see a clear power-law scaling in their spectra, we do not observe such clear scaling in the second order structure function.

\begin{acknowledgments}
We acknowlege the assistance of B. Basel and H. Ligtenberg, and we thank B. Benschop, M. Bos, and G.W. Bruggert for their technical support. This study was financially supported by the Simon Stevin Prize of the Technology Foundation STW of The Netherlands.
\end{acknowledgments}

\end{document}